\documentclass[a4paper,11pt]{article}

\usepackage{contribution}
\newcommand{\weblink}[2][]{%
    \ifthenelse{\equal{#1}{}}%
    {\textnormal{\url{#2}}}%
    {\textnormal{\href{#2}{#1}}}%
}

\newcommand{\acknowledgements}[1]{%
  \bigskip\bigskip
  \textsf{\textbf{\Large Acknowledgements}} \\[2ex]
  {#1}
  \bigskip
}

\def\beq{\begin{equation}}
\def\eeq#1{\label{#1}\end{equation}}
\def\eeqn{\end{equation}}

\def\beqa{\begin{eqnarray}}
\def\eeqa#1{\label{#1}\end{eqnarray}}
\def\eeqan{\end{eqnarray}}

\let\bar=\overbar

\def\etal{{\it et al.}}

\def\Dslash{\not{\hbox{\kern-4pt $D$}}}
\def\dslash{\not{\hbox{\kern-2pt $\del$}}}

\def\msb{{\bar{\ssstyle M \kern -1pt S}}}

\newcommand{\contribution}[7][]{%
  \clearpage
  \thispagestyle{plain}
  \ifthenelse{\equal{#1}{}}
  {\hypersetup{pdftitle={#2}}}
  {\hypersetup{pdftitle={#1}}}
  \hypersetup{pdfauthor={{#3} {#4}}}
  {\centering\normalfont\LARGE\bfseries\sffamily #2 \par\nobreak}
  \lhead{}
  \chead{%
    \textit{\footnotesize XIV International Conference on Hadron Spectroscopy
      (\weblink[\textit{hadron2011}]{http://www.hadron2011.de}), 13-17 June 2011, Munich, Germany}%
  }
  \rhead{}
  \bigskip
  \begin{center}
    {#3} {#4}\ifthenelse{\equal{#6}{}}{}{\footnote{\weblink[#6]{mailto:#6}}}
    \ifthenelse{\equal{#7}{}}{}{#7} \\
    \textit{#5}
  \end{center}
  \bigskip
}

\renewcommand{\abstract}[1]{%
  \begin{center}
    \begin{minipage}{0.85\textwidth}
      \begin{footnotesize}
        #1
      \end{footnotesize}
    \end{minipage}
  \end{center}
  \bigskip
}

\begin{document}

% % % % % % % % % % % % % % % % % % % % % % % % % % % % % % % % % % % % % % % % %
% your proceedings

%%%%%%%%%%%%%%%%%%%%%%%%%%%%%%%%%%%%%%%%%%%%%%%%%%%%%%%%%%%%%%%%%%%%%%%%%%%%%%%%%
%
% template for hadron2011 contribution
%
% please do not rename this file
%
% to create document run
%
%     pdflatex hadron2011.tex
%
%%%%%%%%%%%%%%%%%%%%%%%%%%%%%%%%%%%%%%%%%%%%%%%%%%%%%%%%%%%%%%%%%%%%%%%%%%%%%%%%%
{  % do not remove

%%%%%%%%%%%%%%%%%%%%%%%%%%%%%%%%%%%%%%%%%%%%%%%%%%%%%%%%%%%%%%%%%%%%%%%%%%%%%%%%%
% please define your macros here
%
%%%%%%%%%%%%%%%%%%%%%%%%%%%%%%%%%%%%%%%%%%%%%%%%%%%%%%%%%%%%%%%%%%%%%%%%%%%%%%%%%

%%%%%%%%%%%%%%%%%%%%%%%%%%%%%%%%%%%%%%%%%%%%%%%%%%%%%%%%%%%%%%%%%%%%%%%%%%%%%%%%%
% define title, author, and address
% contribution[short title]{title}{author first name}{author last name}{author address}{author email}{collaboration}
% the short title will appear in the page headers and the TOC of the book of proceedings
% the last two arguments may be left empty
\contribution[COMPASS - a facility to study QCD]  % short title (optional)
{COMPASS - a facility to study QCD}  % title
{Eva-Maria.}{Kabu{\ss}}  % first and last name of author
{Institut f\"ur Kernphysik\\
 Johannes-Gutenberg-Universit\"at Mainz \\
 D-55099 Mainz, Germany}  % author address
{emk@kph.uni-mainz.de}  % author email optional
{on behalf of the COMPASS Collaboration}  % collaboration (optional)
%
%%%%%%%%%%%%%%%%%%%%%%%%%%%%%%%%%%%%%%%%%%%%%%%%%%%%%%%%%%%%%%%%%%%%%%%%%%%%%%%%%

%%%%%%%%%%%%%%%%%%%%%%%%%%%%%%%%%%%%%%%%%%%%%%%%%%%%%%%%%%%%%%%%%%%%%%%%%%%%%%%%%
% abstract
\abstract{%
An overview on the new COMPASS II experimental programme is presented. The main
topics include a study of Primakoff reactions, generalised parton distributions
via deeply virtual Compton scattering and transverse momentum dependent
distributions in Drell-Yan processes in the pion scattering off polarised 
protons.
Moreover, the studies of semi-inclusive deep inelastic scattering on unpolarised target will be continued.
}
%
%%%%%%%%%%%%%%%%%%%%%%%%%%%%%%%%%%%%%%%%%%%%%%%%%%%%%%%%%%%%%%%%%%%%%%%%%%%%%%%%%

%%%%%%%%%%%%%%%%%%%%%%%%%%%%%%%%%%%%%%%%%%%%%%%%%%%%%%%%%%%%%%%%%%%%%%%%%%%%%%%%%
% main text
% for short contributions sections are optional
\section{Introduction}
The COMPASS experiment \cite{emk:COMPASS} uses the unique CERN SPS M2 beamline that is able
to deliver high-energy hadron and polarised muon beams. 
The COMPASS apparatus \cite{emk:spectrometer} consists of a high-resolution two stage forward spectrometer and a versatile target section allowing to use
polarised and unpolarised targets for the various physics programmes.
Up to now 
measurements were performed to study the longitudinal and transverse
spin structure of the nucleon in polarised muon-nucleon scattering as 
well as meson spectroscopy and Primakoff reactions using negatively and
positively charged hadron beams.

Recently, the COMPASS II proposal \cite{emk:COMPASSII} was submitted to
improve the knowledge of the momentum structure of the nucleon towards
a three dimensional picture. For this a series of new measurements is planned.
A study of generalised parton distributions (GPD) will be done in exclusive reactions
like deeply virtual Compton scattering (DVCS) and deeply virtual meson 
production (DVMP) \cite{emk:mueller,emk:rady}. In parallel the study of flavour separation and hadron fragmentation in semi-inclusive deep inelastic scattering (SIDIS) will be
continued. Drell-Yan processes will be used for a complementary study of transverse momentum dependent distributions (TMD) using a transversely polarised target
\cite{emk:arnold}.
At very low momentum transfers Primakoff reactions can be used to extract
pion and kaon polarisabilities.
The COMPASS II proposal  was approved on December 2010 for an initial data 
taking of three years.

\section{Primakoff reactions}
Chiral perturbation theory allows to predict the low energy behaviour of 
Compton scattering off pions and kaons \cite{emk:chiral}. The deviation from 
the response of a pointlike particle
is parametrised in first order by the polarisabilities $\alpha$ and $\beta$ 
in terms of  $\alpha-\beta$ and $\alpha+\beta$. Although a number of 
measurements, mainly for   $\alpha_{\pi}-\beta_{\pi}$, were performed no firm 
conclusion could be drawn on the comparison to the chiral prediction for 
$\alpha_{\pi}-\beta_{\pi}=5.70 \pm 1.0\cdot 10^{-4}$~fm$^3$ \cite{emk:primakoff},

One of the unique features at COMPASS is
the availability of pion and muons beams, where pointlike muons can be used
for comparison and systematic studies. Switching between muon and pion beams
is possible within few hours. Exploratory measurements were performed in 2004
and 2009 to establish the feasibility and to study the achievable systematic
uncertainties in such measurements. With a total measurement time of about
120~d (90~d for $\pi$ and 30~d for $\mu$, planned in 2012) a statistical
precision of   $\alpha_{\pi}-\beta_{\pi}$ of $0.7\cdot 10^{-4}$~fm$^3$
is expected. In addition, $\alpha_{\pi}+\beta_{\pi}$ as well as higher order
terms will be accessible. The small kaon component in the negatively charged 
hadron beam will allow a first measurement of the kaon polarisability.
Reactions with one or two neutral pions instead of a photon in the final state 
will allow to investigate the chiral anomaly and get further insight into 
chiral dynamics. \cite{emk:primakoff}.

\section{GPD measurements}

Generalised parton distributions provide a unified description of form factors
and parton distributions and allow transverse imaging of the nucleon \cite{emk:tomo}. Two of
them, GPD $H$ and $E$, can be studied in DVCS using unpolarised and transversely
polarised targets while all four GPDs, $H$, $E$, $\tilde{H}$ and $\tilde{E}$
can be accessed in DVMP. GPD $H$ and $E$ allow also to study the total
angular momentum carried by quarks inside the nucleon \cite{emk:ji}.

The measurements will exploit the availability of 160 GeV $\mu^{\pm}$ beams
with opposite polarisation 
scattering off a liquid hydrogen target in the initial phase. For a later
stage also a transversely polarised NH$_3$ target is under consideration.
The experiment will cover a unique kinematic range at intermediate values of
the Bjorken scaling variable $x_{\rm Bj}$ 
($0.01<x_{\rm Bj}<0.1)$, a region not yet covered by any other experiment. In this 
kinematic range the
Bethe-Heitler process is a competing process. Being well known it can serve 
as a reference process. GPDs will be studied measuring the so called 
``beam charge
and spin'' difference and sum which allow to extract the real and imaginary 
part of GPD $H$ from the obtained Compton amplitudes. The
$t$ dependence of the DVCS cross section can be parametrised as 
$\frac{{\rm d}\sigma}{{\rm d} t} \sim \exp (-B(x_{\rm Bj}) |t|)$. Figure~\ref{fig:emk_slope}~(left)
shows the projected statistical accuracy using two years of data taking 
for a measurement of the $x_{\rm Bj}$ dependence
of the $t$~slope parameter $B$ of the DVCS cross section for two values of the slope parameter $\alpha'$ with $B(x_{\rm Bj})=B_0 + 2 \alpha' 
\log(x_0/x_{\rm Bj})$. At small $x_{\rm Bj}$,
$B$ can be related to the transverse size of the nucleon 
$<r^2_{\perp}(x_{\rm Bj})> \approx 2 B(x_{\rm Bj})$.

\begin{figure}[htb]
\vspace{-1cm}
  \begin{center}
    % please do not add file name extension this makes switching between latex and pdflatex easier
    \includegraphics[width=0.51\textwidth]{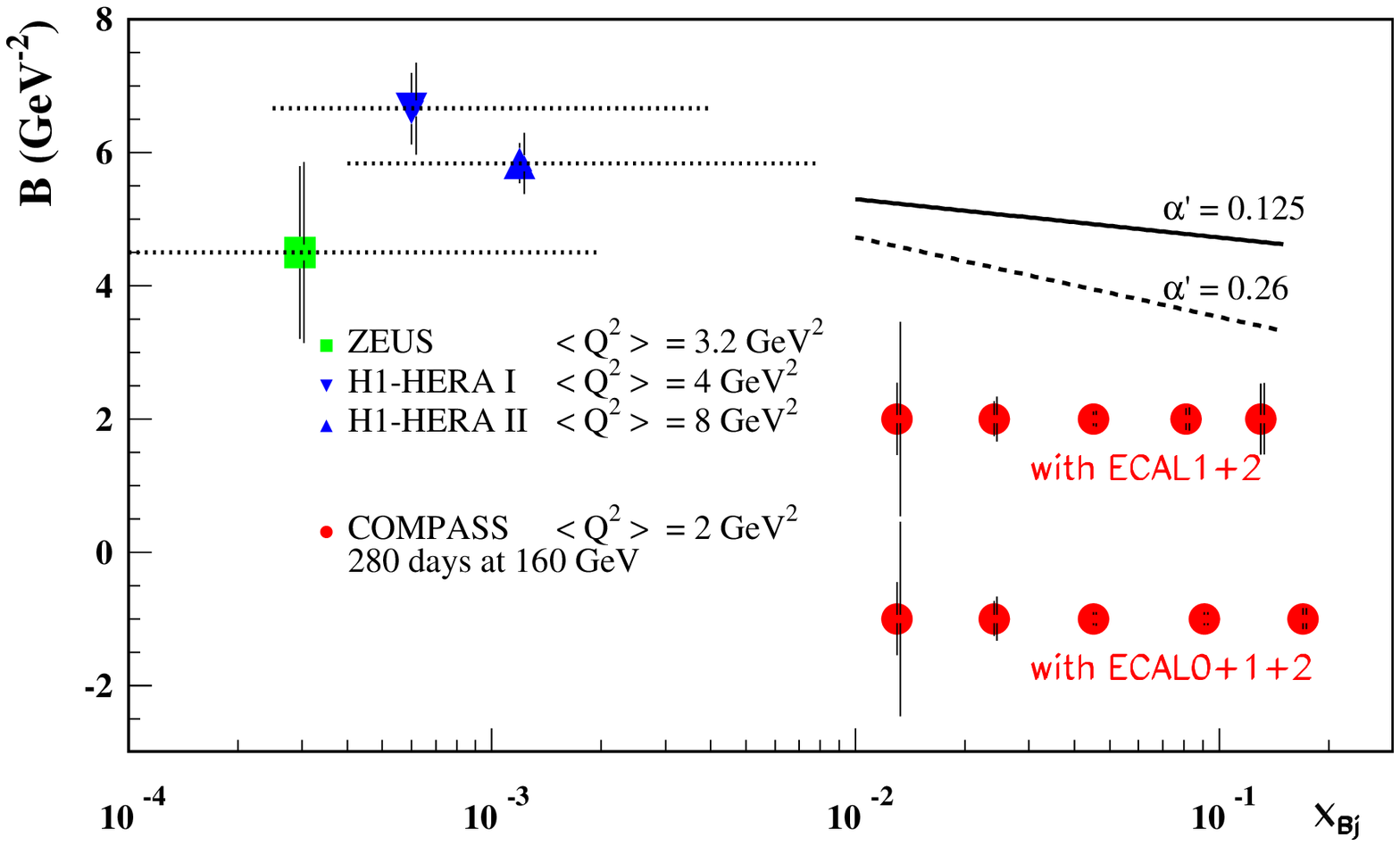}
    \includegraphics[width=0.47\textwidth]{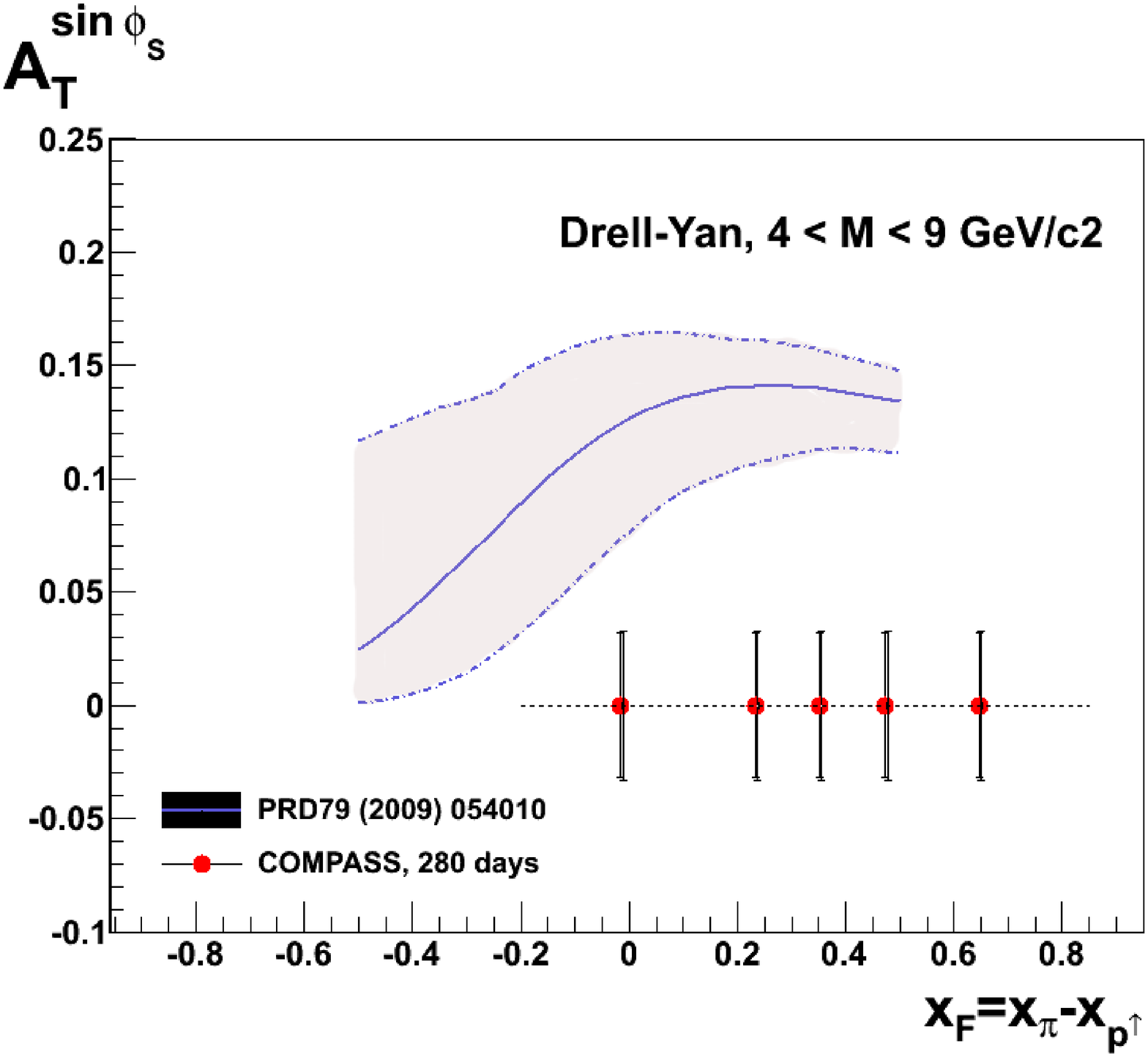}
    \caption{Projections for the statistical accuracy of (left) the slope parameter 
$B(x_{\rm Bj})$ of the DVCS cross section for $1<Q^2/({\rm GeV}/c)^2 < 8$; also shown are published data with a similar $Q^2$, (right) the Sivers asymmetry for two years of data taking compared to the prediction of \cite{emk:anselmino} for the high mass region.}
    \label{fig:emk_slope}
  \end{center}
\end{figure}
 
For these measurements the COMPASS spectrometer has to be upgraded with a 2.5~m long liquid hydrogen target, a 4~m long recoil proton detector and hermetic coverage with electromagnetic calorimetry. The feasibility of the measurement was
studied in several test measurements, the latest in 2009 which showed a clear
DVCS signal at high $x_{\rm Bj}$. A first data taking with the full set-up is planned
for the end of 2012.

In parallel to the DVCS and DVMP measurements the study of SIDIS will be 
continued. The main aim is to improve the knowledge on fragmentation functions,
especially for strange quarks, and to constrain the strange quark distribution. 
Due to the large amount of data, detailed studies of the 
dependence on several kinematic variables like $x_{\rm Bj}$, $z$, $Q^2$ and $p_T$ for
hadron multiplicities will be possible, which will also provide
high precision inputs to NLO pQCD analyses. Moreover, the study of TMDs in 
unpolarised SIDIS off protons will complement the previous measurements with 
the COMPASS polarised $^6$LiD target.

\section{Drell-Yan measurements}
Complementary aspects of the nucleon can be studied with TMDs. They provide
a dynamic picture using the intrinsic  transverse momentum of partons inside
the nucleon. Previously COMPASS had access to e.g. the Sivers TMD and the 
Boer-Mulders TMD in SIDIS off transversely polarised targets. In these 
measurements TMDs are convoluted with fragmentation functions.
An alternative approach is offered by Drell-Yan processes. Here a convolution of
two TMDs from the projectile and the target is studied. Such data will allow
a test of the factorisation ansatz: the sign of the Sivers and Boer-Mulders 
functions are expected to be opposite in DY and SIDIS.

The measurements will be done using a 190 GeV pion beam impinging of the 
transversely polarised COMPASS NH$_3$ target. In the final state a pair of
oppositely charged muons will
be selected using a new dimuon trigger system. To reduce the high hadron 
background an absorber will be placed
downstream of the target which also contains a tungsten plug to absorb the
non-interacting beam. The dominant process is the annihilation of a valence 
anti-quark from the pion with an valence quark from the proton. The 
combinatorial background can be estimated using like-sign muon pairs.
Muon pairs in the mass range of ($4\le M_{\mu\mu}/({\rm MeV}/c^2)\le9$)
will be used to extract the signal as the estimated background is very small
in this region. In two years of data taking 230000 high mass DY events are
expected. In Fig.~\ref{fig:emk_slope}~(right) the achievable statistical accuracy
for the Sivers asymmetry is compared to a recent theoretical prediction \cite{emk:anselmino}.

%%%%%%%%%%%%%%%%%%%%%%%%%%%%%%%%%%%%%%%%%%%%%%%%%%%%%%%%%%%%%%%%%%%%%%%%%%%%%%%%%
% acknowledgements (optional)
\acknowledgements{%
This work was supported by the Bundesministerium f\"ur Bildung und Forschung
(Germany).
 
}

%%%%%%%%%%%%%%%%%%%%%%%%%%%%%%%%%%%%%%%%%%%%%%%%%%%%%%%%%%%%%%%%%%%%%%%%%%%%%%%%%
% bibliographic items can be constructed using the LaTeX format in SPIRES
% see http://www.slac.stanford.edu/spires/hep/latex.html
% SPIRES will also supply the CITATION line information; please include it

%
%%%%%%%%%%%%%%%%%%%%%%%%%%%%%%%%%%%%%%%%%%%%%%%%%%%%%%%%%%%%%%%%%%%%%%%%%%%%%%%%%

}  % do not remove

%%% Local Variables: 
%%% mode: latex
%%% TeX-master: "../hadron2011.tex"
%%% End: 

\end{document}